\documentclass[twocolumn,prl,aps,showpacs,amssymb,superscriptaddress]{revtex4}
\usepackage{bm}
\usepackage{graphicx}
\usepackage{amsmath}

\begin{document}
\title{Electron-Magnon Scattering in Anomalous Hall Effect}

\author{Shengyuan A. Yang}
\affiliation{Department of Physics, The University of Texas, Austin,
Texas 78712, USA}

\author{Hui Pan}
\affiliation{Department of Physics, The University of Texas, Austin,
Texas 78712, USA} \affiliation{Department of Physics, Beijing
University of Aeronautics and Astronautics, Beijing 100083, China}

\author{Wong-Kong Tse}
\affiliation{Department of Physics, The University of Texas, Austin,
Texas 78712, USA}

\author{Qian Niu}
\affiliation{Department of Physics, The University of Texas, Austin,
Texas 78712, USA}

\begin{abstract}
We study the role played by electron-magnon scattering in the
anomalous Hall effect. We find that it has important contributions
distinct from other scattering processes like impurities scattering
and phonon scattering. As a demonstration, we calculate the Hall
conductivity for a two dimensional Dirac model. The result indicates
that as system control parameter varies, the competition between
magnon scattering and other types of scattering changes the Hall
conductivity drastically. In particular, the side jump contribution
could acquire a strong temperature dependence.
\end{abstract}

\date{\today}
 \maketitle

The anomalous Hall effect (AHE), in which a transverse voltage is
induced by a longitudinal current flow in ferromagnetic materials,
is one of the most intriguing effects in physics. While it has been
widely used experimentally as a standard technique for the
characterization of ferromagnets, the theoretical study of the AHE
proves to be complicated and is a subject full of controversial
issues and conflicting results~\cite{naga2009}. In recent years, an
important connection has been established between the AHE and the
Berry phase of Bloch
electrons~\cite{berr1984,sund1999,jung2002,onod2002}. This triggers
revived interest in this subject and is followed by extensive
researches both theoretically and experimentally~\cite{theoexp}. It
is now generally accepted that the AHE is due to spin-orbit coupling
in ferromagnets, and apart from an intrinsic contribution which is
scattering independent, there are also important extrinsic
contributions to the AHE. Especially, there is a peculiar side jump
contribution that arises from scattering, but does not depend on the
scattering strength.

Up to now, most of the theoretical studies of the AHE only take into
account the impurity scattering, despite that many experimental
measurements are performed at finite temperatures hence other
scattering processes like phonon scattering and magnon scattering
should also be relevant. It has been shown that the phonon
scattering produces similar contributions as impurity
scattering~\cite{lyo1973}, which seems to imply that the side jump
contribution should only weakly depend on temperature. However,
recent experiment by Tian \emph{et al.} shows that the side jump
does have strong temperature dependence~\cite{tian2009}.

In this paper, we show that the magnon scattering, which has been
largely overlooked so far, has distinct contributions to the AHE
from both impurity scattering and phonon scattering. The underlying
reason for this difference is that the magnon scattering involves
spin-flip, whereas the impurity scattering and the phonon scattering
are both spin independent. The extrinsic contribution to the AHE
turns out to depend sensitively on the type of spin structure of the
scattering process, which has also been noticed in the recent study
of the magnetic impurity scattering~\cite{miha2008,nunn2008}.
Therefore, the anomalous Hall conductivity would change drastically
as two or more types of scattering compete. In particular, the side
jump contribution will depend on the relative scattering strength of
different types of scattering processes, hence should have strong
temperature dependence.

We begin with a description of the electron-magnon scattering
process in ferromagnetic systems. The exchange coupling between
conduction electron and local magnetic order parameter can be
written as
\begin{equation}\begin{split}\label{interact}
\hat{H}_\text{int}&=-J\int d\bm{r}\left[
\hat{\bm{\sigma}}(\bm{r})\cdot\bm{S}(\bm{r})\right]\\
&=-\frac{J}{2} \int
d\bm{r}\left(\hat{\sigma}_+S_-+\hat{\sigma}_-S_++2\hat{\sigma}_zS_z\right),
\end{split}
\end{equation}
where $J$ is the exchange coupling constant, $\hat{\bm{\sigma}}$ is
the vector of Pauli matrices for conduction electron spin, $\bm{S}$
is the local spin, $\hat{\sigma}_\pm\equiv \hat{\sigma}_x\pm
i\hat{\sigma}_y$, $S_\pm\equiv S_x\pm iS_y$, and hat means the
quantity is a matrix in spin space. The last term above describes
the exchange splitting which should be included in the
non-interacting part of the Hamiltonian, whereas the first two terms
describe the electron-magnon scattering. Using Holstein-Primakoff
representation, for temperatures below Curie point when the number
of local spin flips is much smaller than the total spin, we can
write the interaction Hamiltonian as
\begin{equation}\label{emag}
\hat{H}_\text{int}=-\frac{J\sqrt{2S}}{2}\frac{1}{V^2}
\sum_{\bm{k,q}}\left(c^\dagger_{\bm{k}+\bm{q}\uparrow}c_{\bm{k}\downarrow}
a^\dagger_{-\bm{q}}+
c^\dagger_{\bm{k}+\bm{q}\downarrow}c_{\bm{k}\uparrow}
a_{\bm{q}}\right),
\end{equation}
where $V$ is the system volume, $c^\dagger$ ($c$) and $a^\dagger$
($a$) are the electron and magnon creation (annihilation) operators
respectively. From Eq.(\ref{emag}), it is clear that the magnon
scattering flips spin, unlike impurity scattering and phonon
scattering which are spin independent.

When the energy of magnons involved in the scattering is much lower
than the Fermi energy and the electron energy spectrum (and density
of states) varies smoothly around Fermi surface, we can approximate
the scattering process as quasi-elastic. In typical ferromagnetic
materials like Fe or Co, the magnon has a large effective mass about
$10^{-29}\sim10^{-28}$kg~\cite{magmass}. The quasi-elastic treatment
will be a good approximation if the electron effective mass at Fermi
level is much smaller than that of the magnon. In this case, the
electron sees an effective scattering potential
\begin{equation}\label{magv}
\hat{V}_\text{m}(\bm{q})=\frac{1}{\sqrt{2}}V^o_\text{m}(q)(\hat{\sigma}_++\hat{\sigma}_-),
\end{equation}
where $V^o_\text{m}(q)=-J\sqrt{Sn_\text{m}(q)}/2$ is the orbital
part of the scattering potential, $n_\text{m}(q)$ is the
distribution function of magnon, and the spin part
$\hat{\sigma}_\pm$ is off-diagonal representing spin-flip processes.

To demonstrate that the magnon scattering gives distinct
contributions to the AHE, we calculate the Hall conductivity of the
two dimensional (2D) Dirac model. The AHE from impurity scattering
in this model has been studied previously by Sinitsyn \emph{et
al.}~\cite{sini2007prb}. We choose this model not only because of
its simplicity to demonstrate our ideas, but also because it
describes low energy physics of interesting systems such as
graphene~\cite{cast2009}, kagome lattice~\cite{guo2009} and surface
states of topological insulator~\cite{qi2008} (though it should be
noticed that for graphene and kagome lattice the `spin' refers to
the sublattice degrees of freedom rather than electron spin as
discussed here). Therefore the results presented here will also be
important in understanding transport properties of these systems.

The AHE occurs in 2D Dirac model when a suitable symmetry breaking
mechanism is introduced. Our model Hamiltonian reads (we set
$\hbar=1$ in the following)
\begin{equation}\label{dirac}
\hat{\mathcal{H}}=v(k_x\hat{\sigma}_x+k_y\hat{\sigma}_y)+\Delta
\hat{\sigma}_z,
\end{equation}
where the last term is the symmetry breaking term which introduces a
gap of $2\Delta$. We assume that this model describes low energy
physics near the Fermi surface of certain 2D ferromagnetic system,
hence the term $\Delta \hat{\sigma}_z$ represents the exchange
splitting which corresponds to the last term in Eq.(\ref{interact}).

The eigenstates of the system are given by
$\psi_{\bm{k}}^\text{c,v}(\bm{r})=
(1/\sqrt{V})e^{i\bm{k}\cdot\bm{r}}|u_{\bm{k}}^\text{c,v}\rangle$
with energy eigenvalues
$\varepsilon^\text{c,v}(\bm{k})=\pm\sqrt{(vk)^2+\Delta^2}$, where c
and v stand for conduction and valence band respectively, and
$|u_{\bm{k}}^\text{c,v}\rangle$ is the spin part of the eigenstate
which can be written as
\begin{equation}
|u_{\bm{k}}^\text{c}\rangle=\left(
                       \begin{array}{c}
                         \cos\frac{\theta}{2} \\
                         \sin\frac{\theta}{2}e^{i\phi} \\
                       \end{array}
                     \right),
\qquad |u^\text{v}_{\bm{k}}\rangle=\left(
                       \begin{array}{c}
                         \sin\frac{\theta}{2} \\
                         -\cos\frac{\theta}{2}e^{i\phi} \\
                       \end{array}
                     \right),
\end{equation}
with $\theta$ and $\phi$ being the spherical angles of the vector
$(vk_x,vk_y,\Delta)$ such that
$\cos\theta=\Delta/\sqrt{(vk)^2+\Delta^2}$ and $\tan\phi=k_y/k_x$.
Due to spin-orbit coupling, the eigen-spinors are
$\bm{k}$-dependent.

We evaluate the Hall conductivity by using the Kubo-Streda
formalism~\cite{stre1982,crep2001}. In this approach, the Hall
conductivity can be separated into two parts,
$\sigma_{xy}=\sigma_{xy}^\text{I}+\sigma_{xy}^\text{II}$, where
$\sigma_{xy}^\text{I}$ is a Fermi surface contribution, and
$\sigma_{xy}^\text{II}$ is a Fermi sea contribution for which we
only need to retain the scattering-free component in the weak
disorder limit~\cite{sini2007prb}. All the important scattering
effects are contained in $\sigma_{xy}^\text{I}$, which takes the
form
\begin{equation}
\sigma_{xy}^\text{I}=\frac{e^2}{2\pi V}\text{Tr}\left\langle
\hat{v}_x
\hat{G}^R(\varepsilon_F)\hat{v}_y\hat{G}^A(\varepsilon_F)\right\rangle,
\end{equation}
where $\hat{G}^R$ and $\hat{G}^A$ are the retarded and advanced
Green's functions respectively, $\hat{v}_{x,y}=v\hat{\sigma}_{x,y}$
are the velocity operators, $\varepsilon_F$ is the Fermi energy, the
trace is taken over both momentum and spin spaces and the bracket
means statistical average over disorder configurations. In the
following calculation, we take the Fermi energy to be in the
conduction band.

The intrinsic contribution results from the Berry curvatures of
spin-orbit coupled bands and is independent of scattering. In
Kubo-Streda formalism, the intrinsic contribution comes from the
scattering-free components of $\sigma_{xy}^\text{I}$ and
$\sigma_{xy}^\text{II}$~\cite{sini2007prb}. It can be decomposed
into two parts
$\sigma^\text{int}_{xy}=\sigma^\text{int(v)}_{xy}+\sigma^\text{int(c)}_{xy}$,
where $\sigma^\text{int(v)}_{xy}$ is the contribution from all the
completely occupied valence bands below the Fermi surface and
$\sigma^\text{int(c)}_{xy}$ is the contribution from the partially
filled conduction band where the Fermi surface lies in. The
contribution from completely filled bands
$\sigma^\text{int(v)}_{xy}$ has a topologically quantized value
$Ne^2/(2\pi)$ with $N$ being an integer known as the first Chern
number. The calculation of $N$ goes beyond any low energy effective
model since it involves the entire Fermi sea. On the contrary, the
contribution $\sigma^\text{int(c)}_{xy}$ from the partially filled
conduction band can be regarded as a Fermi surface
property~\cite{hald2004}. For Dirac model, we have
\begin{equation}\label{int}
\sigma^\text{int(c)}_{xy}=\frac{e^2}{4\pi}(1-\cos\theta_F),
\end{equation}
where $\theta_F$ is the spherical angle $\theta$ evaluated on the
Fermi surface when $k=k_F$.

The extrinsic contribution consists of the side jump and the skew
scattering. In the semiclassical picture, the side jump arises from
the coordinate shift of a wave-packet during the scattering process,
and the skew scattering appears due to the asymmetry of scattering
rate for higher order scattering processes~\cite{naga2009}. It has
been clarified recently that the skew scattering contribution as
defined in the semiclassical picture actually contains two different
parts: a conventional skew scattering part with $n^{-1}$ dependence
and an intrinsic skew scattering part with $n^0$ dependence with $n$
being the disorder density~\cite{sini2008}. According to its
parametric dependence, the intrinsic skew scattering can be included
as part of the side jump. In Kubo-Streda formalism, the various
contributions listed above have been identified with different sets
of Feynman diagrams in the self-consistent non-crossing
approximation~\cite{sini2007prb}.

Let's first consider a clean system with only magnon scattering. In
the quasi-elastic approximation, the disorder lines in Feynman
diagrams do not carry energy arguments. Since the population of
magnon bath is conserved in steady state, each disorder line must
have a pair of $\hat{\sigma}_+$ and $\hat{\sigma}_-$ at the two
ends, corresponding to magnon emission and absorption processes.
Therefore the conventional skew scattering which involves third
order scattering events must vanish. Furthermore, due to the angular
average at velocity vertices, the intrinsic skew scattering also
vanishes, leaving only the side jump contribution,
\begin{equation}\label{mext}
\sigma^\text{ext}_{xy}=\frac{e^2}{4\pi}\cos\theta_F,
\end{equation}
which cancels with the part of intrinsic contribution that depends
on Fermi energy, such that the final result becomes a constant value
$e^2/(4\pi)$ and is the same as the intrinsic contribution for a
completely filled conduction band, i.e. in the limit
$\theta_F\rightarrow \pi/2$~\cite{quanex}.

Next we shall include the impurity scattering as well and
investigate the competition between magnon scattering and impurity
scattering in the AHE. For simplicity, we consider the short range
impurity as been studied in Ref.~\cite{sini2007prb}. The result of
the total Hall conductivity is (including only
$\sigma_{xy}^\text{int(c)}$ for intrinsic contribution)
\begin{widetext}
\begin{equation}\label{sig}
\sigma_{xy}=\frac{e^2}{4\pi}(1-\cos\theta_F)-\frac{e^2}{\pi}\frac{\sin^2\theta_F\cos\theta_F(1-\zeta)}{
(1+3\cos^2\theta_F)+4\sin^2\theta_F\zeta}
-\frac{e^2}{\pi}\frac{\sin^4\theta_F\cos\theta_F(\frac{3}{4}-\zeta+2\eta)}{\left[
(1+3\cos^2\theta_F)+4\sin^2\theta_F\zeta\right]^2},
\end{equation}
\end{widetext}
where $\zeta=\tau_\text{m}^{-1}/\tau_\text{i}^{-1}$ is the ratio of
magnon scattering rate to impurity scattering rate with
\begin{equation}
\tau_\text{m,i}^{-1}=2\pi
\int\frac{d^2\bm{k}'}{(2\pi)^2}\left\langle
\left|V^o_\text{m,i}(\bm{k}',\bm{k})\right|^2\right\rangle[1-\cos(\phi-\phi')]
\delta(\varepsilon_F-\varepsilon^\text{c}_{\bm{k}'}),
\end{equation}
and $\eta=\tau_{sk}^{-2}/\tau_\text{i}^{-2}$ represents the
conventional skew scattering contribution from impurities. For
standard white noise impurity model, the third order correlation of
scattering potential is zero, so $\eta$ vanishes identically. It
will be nonzero if a non-Gaussian part $V_1$ of the impurity
potential is included~\cite{sini2007prb}, then
$\tau_{sk}^{-2}=n_\text{i}\varepsilon_F^3V_1^3/(4\pi v^4)$. Observe
that when the magnon scattering is dominant over impurity
scattering, i.e. for very large $\zeta$, we recover the result of
Eqs.(\ref{int},\ref{mext}) with a constant value $e^2/(4\pi)$. In
the opposite limit, when impurity scattering dominates,
$\zeta\rightarrow 0$, we retain the previous result by Sinitsyn
\emph{et al}.~\cite{sini2007prb}. From our result Eq.(\ref{sig}), it
is clear that as $\zeta$ varies, which results from the competition
between different scattering mechanisms, the extrinsic contributions
to the Hall conductivity varies drastically and can have a sign
change.

\begin{figure}
\includegraphics[width=0.95\columnwidth]{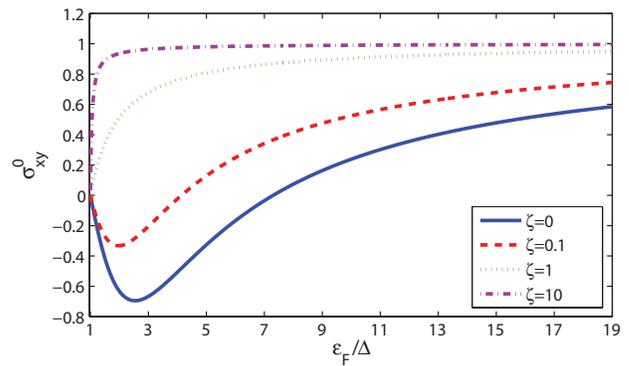}
\caption{\label{fig:sigvsef} (color online). $\sigma^0_{xy}$ plotted
as a function of the Fermi energy $\varepsilon_F$ for fixed values
of $\zeta$. $\sigma^0_{xy}$ is measured in units of $e^2/(4\pi)$,
and $\varepsilon_F$ is measured in units of $\Delta$ which is half
of the gap size.}
\end{figure}

\begin{figure}
\includegraphics[width=0.95\columnwidth]{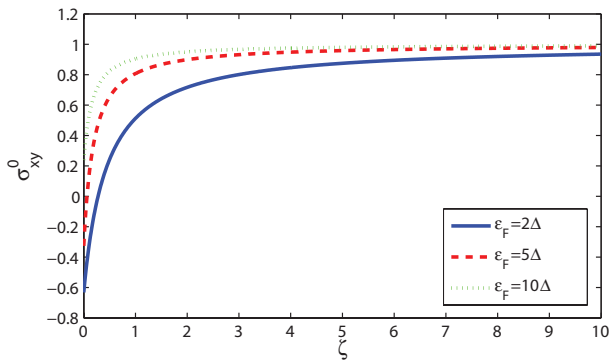}
\caption{\label{fig:sigvszeta} (color online). $\sigma^0_{xy}$
versus the ratio of scattering rates $\zeta$ for fixed values of
Fermi energy $\varepsilon_F$. $\sigma^0_{xy}$ is measured in units
of $e^2/(4\pi)$. The plot shows the crossover from impurity
scattering dominated regime to magnon scattering dominated regime as
$\zeta$ increases.}
\end{figure}

It is of fundamental importance to experimentally separate
contributions from different mechanisms, especially the intrinsic
contribution~\cite{tian2009}. Let us collect the terms of
$\sigma_{xy}$ that are of order $n^0$, denoted as $\sigma_{xy}^0$.
These include the intrinsic contribution and the side jump
(including intrinsic skew scattering). For soft magnon modes, the
ratio $\zeta$ does not sensitively depend on the Fermi energy. In
Fig.\ref{fig:sigvsef}, we plot $\sigma_{xy}^0$ as a function of
Fermi energy for different values of $\zeta$. As $\zeta$ increases,
the curve of Hall conductivity is shifting upward from the impurity
scattering dominated situation and approaching the limiting value
$e^2/4\pi$ for the magnon scattering dominated case. This
competition behavior is more clearly observed in
Fig.\ref{fig:sigvszeta}, where $\sigma_{xy}^0$ is plotted at fixed
Fermi level as a function of $\zeta$. As $\zeta$ increases,
$\sigma_{xy}^0$ increases monotonically. For impurity scattering
dominated case, $\sigma_{xy}^0$ takes negative value for Fermi
energies below $\varepsilon_F\approx 7.3\Delta$. Hence in this
energy range, there is a sign change of $\sigma_{xy}^0$ as $\zeta$
increases, i.e. when magnon scattering gradually takes dominant
place.

As observed from this model calculation, the magnon scattering
indeed plays a quite different role as compared with impurity
scattering. This difference comes from their different structures in
spin space. Magnon scattering flips spin hence its the matrix
element is off-diagonal in spin space while both impurity scattering
and phonon scattering is proportional to the identity in spin space.
As a result, the extrinsic contributions for each type of scattering
involve different combinations of structure factors (such as
$\cos\theta_F$ and $\sin\theta_F$ in our example), which leads to
the competition behavior. Magnetic impurity scattering has yet
another spin structure different from the above three, as being
proportional to $\hat{\sigma}_z$ if the average magnetization is
along $z$-direction. Previous studies indeed show that the magnetic
impurity scattering behaves differently from the normal impurity
scattering for the AHE~\cite{miha2008,nunn2008}. The above analysis
suggests that we could classify various scattering processes
according to their structures in spin space which is the major
factor that determines their contributions to the AHE~\cite{yang}.
The competition between different classes could change the Hall
conductivity dramatically as system control parameter such as
temperature varies.

Finally, we point out that the above discussion is not limited to
ferromagnetic systems with electron spin degrees of freedom. Any
quasi-particle index which has two degrees of freedom can be
generally referred to as `spin' (or pseudospin). Anomalous Hall
effect will arise if the system has `spin'-orbit coupling as well as
`spin' splitting. For example, in a bipartite lattice such as
graphene, the sublattice degree of freedom can be treated as
pseudospin. Anomalous Hall transport occurs in graphene when there
is sublattice symmetry breaking in the system~\cite{xiao2007}. As
another example, for bilayer systems, it is the layer index that
plays the role of pseudospin and the pseudospin splitting can be
realized by imposing a bias between the two layers. In general, our
result indicates that a careful analysis of various scattering
processes according to its pseudospin structure is indispensable in
the study of AHE for these systems.

In summary, we have shown that the electron-magnon scattering plays
an important role in the anomalous Hall effect. The competition
between magnon scattering and other scatterings changes the
anomalous Hall conductivity drastically as system control parameters
are varied. As a result, the side jump contribution can have strong
temperature dependence.

The authors thank A. H. MacDonald and Y. Yao for valuable
discussions. This work is supported by DOE (DE-FG03-02ER45985),
NSF(DMR0906025), Welch Foundation (F-1255), and NSFC (10740420252).


\begin{thebibliography}{99}
\bibitem{quanex} Interestingly, we find that
this cancelation occurs for a generic class of Hamiltonian
$\hat{\mathcal{H}}=\gamma k^n(\cos n\phi_{\bm{k}}\hat{\sigma}_x+\sin
n\phi_{\bm{k}}\hat{\sigma}_y)+\Delta\hat{\sigma}_z$ with integer
winding number $n$. Whether such a behavior is universal or model
dependent is currently under investigation.
\end{thebibliography}
\end{document}